\documentclass[a4paper]{jpconf}
\usepackage{graphicx}
\begin{document}
\title{The elliptic flow of di-electrons in$\sqrt{s_{NN}}$=200 GeV Au+Au collisions at STAR}

\author{Xiangli Cui for the STAR Collaboration}

\address{University of Science and Technology of China, Hefei, 230026, China

 Brookhaven National Laboratory, Upton, NY, 11973, USA
}

\ead{cuixl@rcf.rhic.bnl.gov}

\begin{abstract}
In minimum bias (0-80\%) Au+Au collisions at $\sqrt{s_{NN}}$=200 GeV, the elliptic flow ($v_{2}$) measurements for di-electrons from the low
to intermediate mass region are presented. The differential elliptic flow as a function of transverse momentum in different mass regions are also reported. The simulated and measured $v_{2}$ of di-electrons are compared in the mass regions of $M_{ee}<0.14$ GeV/$c^{2}$ and $0.14<M_{ee}<0.3$ GeV/$c^{2}$.

\end{abstract}

\section{Introduction}

Ultra-relativistic heavy ion collisions provide a unique environment to study the properties of strongly interacting matter at high temperature and high energy density \cite{bibitem1}. One of the crucial probes of this strongly interacting matter is di-lepton measurement. Di-leptons are not affected by the strong interaction once produced. In the low mass region of produced lepton pairs ($M_{ll}<1.1$ GeV/$c^{2}$), we can study the vector meson in-medium properties through their di-lepton decays, where any modifications observed may relate to the possibility of chiral symmetry restoration. In the intermediate mass region ($1.1<M_{ll}<3.0$ GeV/$c^{2}$), the di-lepton spectra are directly related to the thermal radiation of Quark-Gluon Plasma (QGP) \cite{bibitem2,bibitem3}. However, other sources significantly contribute in this mass region, such as $c\overline{c}$ $\rightarrow$ $l^{+}l^{-}X$ or $bb$ $\rightarrow$ $l^{+}l^{-}X$ \cite{bibitem4}.

 Elliptic flow is generated in the early stage of heavy ion collisions, via the transformation of the initial spatial eccentricity of the nuclear overlap region into momentum anisotropies through the action of azimuthally anisotropic pressure gradients \cite{bibitem5}. Di-leptons, which escape from the expanding fireball without reinteraction, will be able to probe specifically this early stage where the elliptic flow first develops. It has been proposed that transverse momentum ($p_{T}$) and $M_{ll}$ dependence of elliptic flow of di-leptons could provide very rich information on specific stages of the fireball expansion, distinguishing partonic and hadronic radiation sources \cite{bibitem5}.

At STAR, the newly installed Time-of-Flight detector (TOF) which covers $|$$\eta$$|$$<0.9$ and with complete azimuthal symmetry ($\bigtriangleup \phi$ = 2$\pi$), offers large acceptance and high efficiency particle identification \cite{bibitem6}. The TOF, combined with the measurements of ionization energy loss ($dE/dx$) from the Time Projection Chamber (TPC) \cite{bibitem7,bibitem8,bibitem9}, enables the electron identification with high purity from low to intermediate $p_{T}$ \cite{bibitem9,bibitem10,bibitem11,bibitem13}. In this article, we present the $M_{ee}$, $p_{T}$, and centrality dependence of di-electron elliptic flow in $\sqrt{s_{NN}}$=200 GeV Au+Au collisions at STAR.
\section{Data analysis and results}

The TPC is the primary tracking device of STAR, and provides the $dE/dx$, momentum, and path-length measurements of particles produced in the collision. The $dE/dx$ is used for particle identification \cite{bibitem8,bibitem9}. The TOF measures the particles' time-of-flight from their decay vertex to a matched hit in the TOF detector. Combined with path-length measurement, the particle velocity ($\beta$) extends the identification of $\pi$ and $K$ up to 1.6 GeV/$c$ and the proton to 3 GeV/$c$ in momentum \cite{bibitem10,bibitem11}. We utilize 220 million minimum-bias (0-80\%) Au+Au collisions taken in year 2010 with full TOF system coverage. With the $\beta$ and $p_{T}$ dependent $dE/dx$ cuts for $p_{T}>0.2$ GeV/$c$ and $|$$\eta$$|$$<1$, electrons can be clearly identified from low to intermediate $p_{T}$. The purity of electron candidates is about 97\% in minimum-bias Au+Au collisions.

The $e^{+}$ and $e^{-}$ candidates from the same events are combined to reconstruct the invariant mass distributions ($M_{ee}$) marked as unlike-sign distributions, which contain both signal and background. The background contains random combinatorial, and other correlated pairs. We use two methods to estimate the background: same-event like-sign, mixed-event unlike-sign techniques. In the same-event like-sign technique, electron candidates from the same event with same charge sign are combined. In the mixed-event technique, unlike-sign electron candidates from different events are combined. In the mixed-event technique, the event used to mix must be in the same centrality bin, vertex $z$-coordinate bin and event plane angle bin. We divide the centrality into 9 bins, vertex-$z$ into 10 bins, and event plane angle into 100 bins so that the events used to mix have a similar structure. In the mass region $M_{ee}<0.7$ GeV/$c^{2}$, we subtract the like-sign background. In the mass region $M_{ee}>0.7$ GeV/$c^{2}$, we subtract the unlike-sign mixed-event background. Because the like-sign and mixed-event background distributions are consistent for the latter mass region, we use mixed-event background for better statistics.

\begin{figure}[h]
\centering
%\%begin{minipage}{32pc}
\includegraphics[width=33pc]{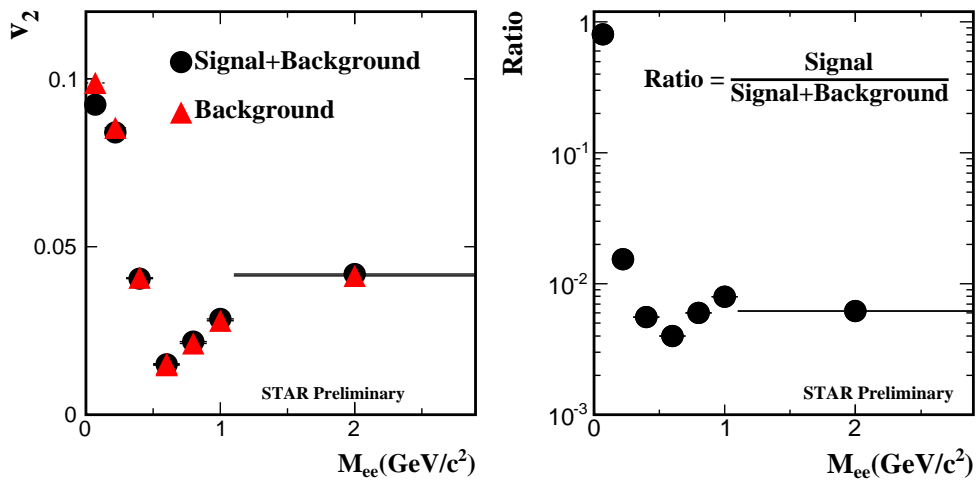}
\caption{\label{label}(left panel) The di-electron $v_{2}$ as a function of $M_{ee}$ for Au+Au $\sqrt{s_{NN}}$=200 GeV collisions. The circles represent the elliptic flow of unlike-sign pairs $( v_{2}^{T})$. The triangles represent the elliptic flow of background pairs $(v_{2}^{B})$. (right panel) Ratio = $\frac{N_{S}}{N_{S+B}}$ as a function of $M_{ee}$ for Au+Au $\sqrt{s_{NN}}$=200 GeV collisions.}
%\end{minipage}%\hspace{3pc}
\end{figure}

We use the event-plane method to calculate the elliptic flow of the di-electron signal. The event-plane is reconstructed using tracks from the TPC. The details of the method are in Refs. \cite{bibitem14,bibitem15}. The di-electron signal $v_{2}$ can be obtained by using the following two formulas:
\begin{eqnarray}
v_{2}^{S}\times\frac{N_{S}}{N_{S+B}} &=& v_{2}^{T}-v_{2}^{B}\times(1-\frac{N_{S}}{N_{S+B}}) ,
\end{eqnarray}
\begin{eqnarray}
v_{2} &=& \langle \cos(2(\phi_{i}-\psi_{2}))/r_{j} \rangle ,
\end{eqnarray}
where $v_{2}^{T}$ is the elliptic flow of the signal+background, $v_{2}^{S}$ is the di-electron signal flow, $v_{2}^{B}$ is the background flow, and we use the formula (2) to calculate $v_{2}^{T}$ and $v_{2}^{B}$ which are shown in the left panel of Fig. 1. $\frac{N_{S}}{N_{S+B}}$ is the ratio of signal over signal+background shown in the right panel of Fig. 1. We use the sub-event method to calculate $r_{j}$ which is resolution of event-plane in centrality $j$, $\phi_{i}$ is the di-electron pair angle, $\psi_{2}$ is the event-plane angle. Angle brackets denote averages over all di-electron pairs in all events. In each mass bin, with the $v_{2}^{T}$, $v_{2}^{B}$, and $\frac{N_{B}}{N_{S+B}}$, we calculate the $v_{2}^{S}$ by the formula (1).

\begin{figure}[h]
\centering
%\begin{minipage}{37pc}
\includegraphics[width=16pc]{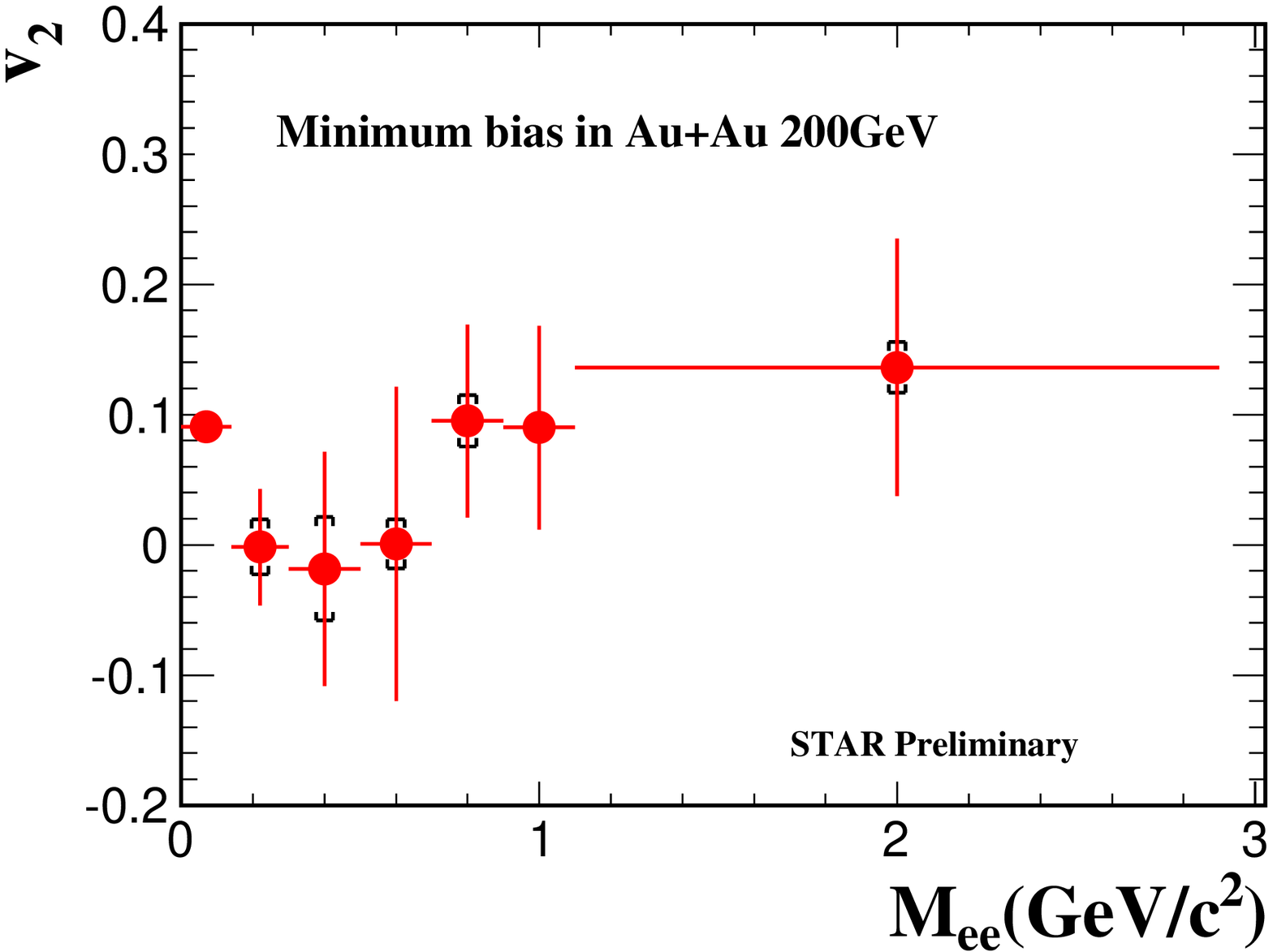}
\includegraphics[width=16pc]{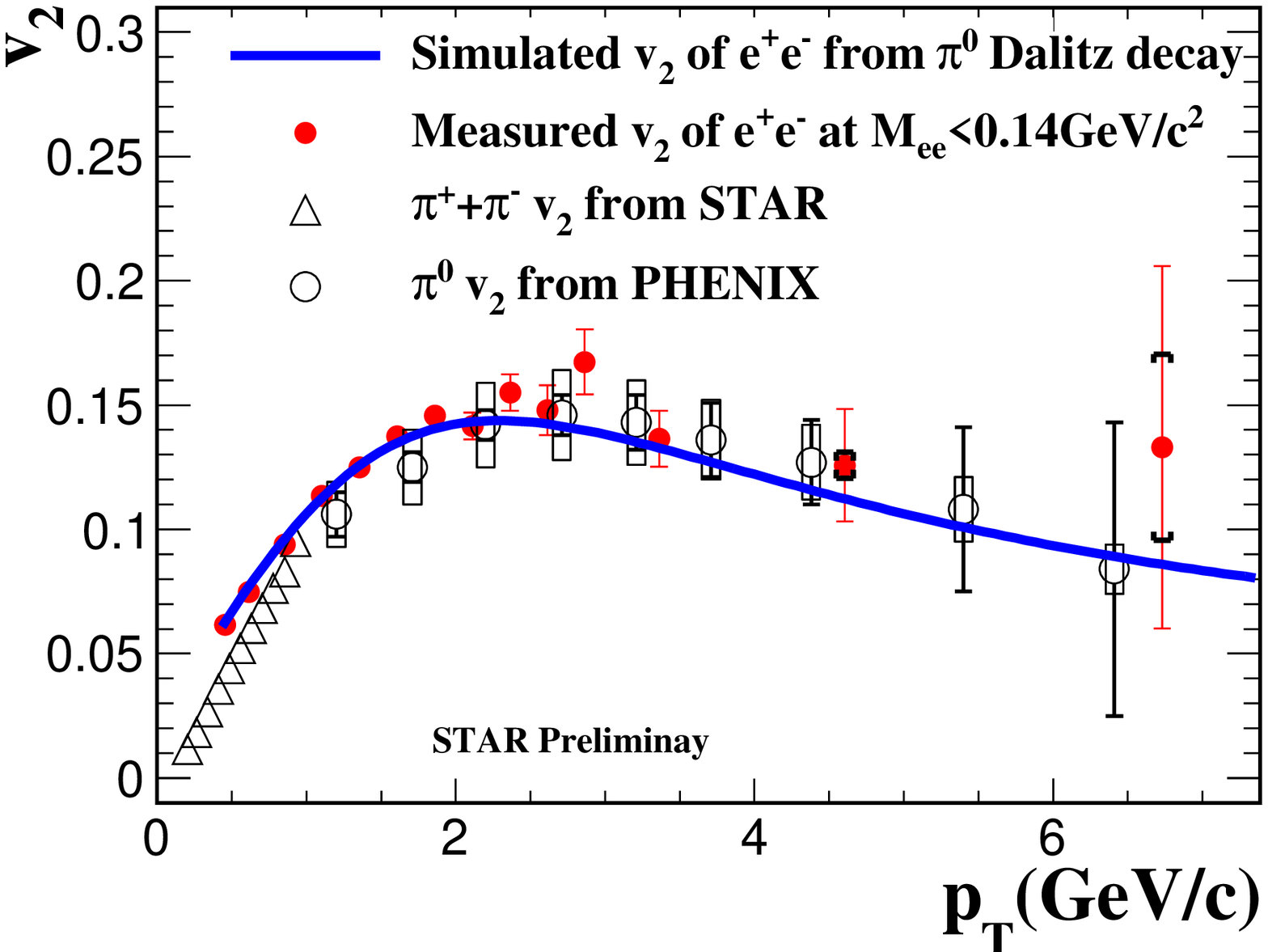}
\includegraphics[width=16pc]{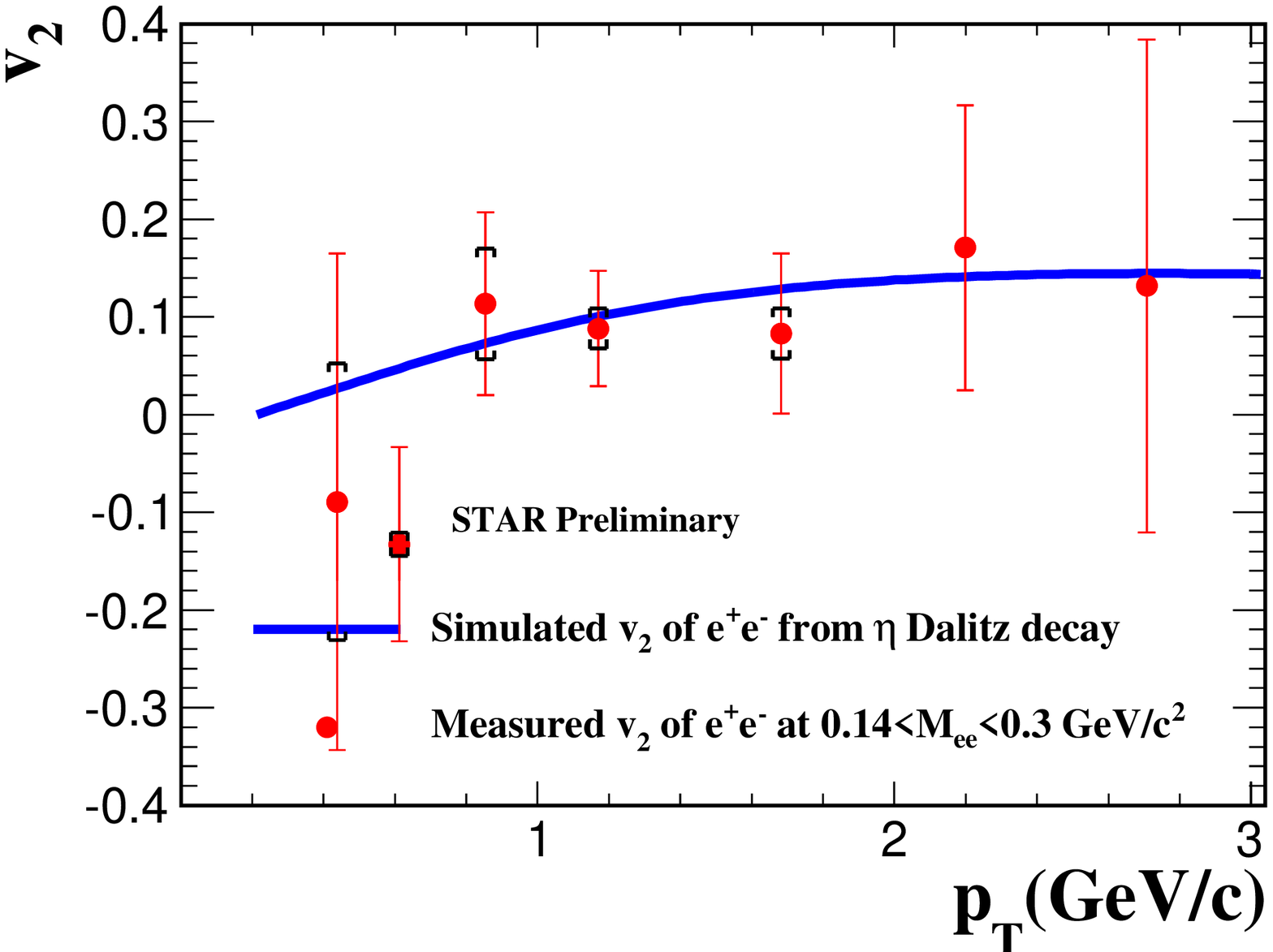}
\caption{\label{label}(upper left panel) The di-electron $v_{2}$ as a function of $M_{ee}$ in minimum bias Au+Au collisions at $\sqrt{s_{NN}}$=200 GeV. (upper right panel) The $v_{2}$ of di-electron as a function of $p_{T}$ (solid symbol) for $M_{ee}<0.14$ GeV/$c^{2}$ in minimum bias Au+Au collisions at $\sqrt{s_{NN}}$=200 GeV. Also shown are the charged pion $v_{2}$ \cite{bibitem16}, neutral pion $v_{2}$ \cite{bibitem17}, and the expected $v_{2}$ (solid curve) of di-electrons from $\pi^{0}$ Dalitz decay. (bottom panel) The $v_{2}$ of di-electron as a function of $p_{T}$ between $0.14<M_{ee}<0.3$ GeV/$c^{2}$ in minimum bias Au+Au collisions at $\sqrt{s_{NN}}$=200 GeV. The expected $v_{2}$ (solid curve) of di-electrons from $\eta$ Dalitz decays is also shown. The error bars and boxes represent statistical and systematic uncertainties, respectively. The systematic errors are estimated by changing track quality cuts.}
%\end{minipage}
\end{figure}

%In minimum bias Au+Au collisions at $\sqrt{s_{NN}}$ = 200GeV, we report the $v_{2}$ of di-electron signal as a function of $M_{ee}$ which are shown in upper left panel of Fig.2. The $p_{T}$ dependent $v_{2}$ of di-electron signal in the mass regions of $M_{ee}<$0.14 GeV/$c^{2}$ and 0.14$<M_{ee}<$0.3 GeV/$c^{2}$ are shown in upper right and bottom panel of Fig.2, respectively.
Figure 2 shows the di-electron $v_{2}$ as a function of $M_{ee}$ (top left panel), $v_{2}$ as a function of $p_{T}$ in the mass region of $M_{ee}<0.14$ GeV/$c^{2}$ (top right panel) and $0.14<M_{ee}<0.3$ GeV/$c^{2}$(bottom panel) in minimum bias Au+Au collisions at $\sqrt{s_{NN}}$=200 GeV. The dominant contributions to di-electron signal for $M_{ee}<0.14$ GeV/$c^{2}$ and $0.14<M_{ee}<0.3$ GeV/$c^{2}$ are $\pi^{0}$ Dalitz decays and $\eta$ Dalitz decays, respectively. We parameterize $\pi^{0}$ $v_{2}$ from low to high $p_{T}$ \cite{bibitem16,bibitem17}, simulate the $\pi^{0}$ Dalitz decay, and obtain the expected di-electron $v_{2}$ from $\pi^{0}$ Dalitz decay shown by the solid curve. The simulated di-electron $v_{2}$ is consistent with the measured $v_{2}$ for $M_{ee}<0.14$ GeV/$c^{2}$. Between $0.14<M_{ee}<0.3$ GeV/$c^{2}$, we assume that the $\eta$ has the same $v_{2}$ as $K_{S}^{0}$ \cite{bibitem15} because the value of their mass are nearly the same, and repeat the same exercise. The simulated $v_{2}$ of di-electrons from $\eta$ Dalitz decay, shown by the solid curve, is consistent with the measured $v_{2}$ for $0.14<M_{ee}<0.3$ GeV/$c^{2}$. The consistency between the expectations and measurements demonstrates the credibility of our method to obtain the di-electron $v_{2}$.

 For $0.5<M_{ee}<0.7$ GeV/$c^{2}$, charm correlation and in-medium $\rho$ contribution might be dominant, and the $p_{T}$ dependent $v_{2}$ of the di-electron signal are shown in the upper left panel of Fig. 3. The upper right panel of Fig. 3 shows the $p_{T}$ dependent $v_{2}$ of the di-electron signal in the mass region of $0.76<M_{ee}<0.8$ GeV/$c^{2}$ and $0.98<M_{ee}<1.06$ GeV/$c^{2}$. At $0.98<M_{ee}<1.06$ GeV/$c^{2}$, the di-electron signal $v_{2}$ is consistent with the measured $v_{2}$ of $\phi$ meson through the $K^{+}K^{-}$ decays \cite{bibitem19}. The bottom panel of Fig. 3 shows the centrality dependence of $v_{2}$ for the mass region $M_{ee}<0.14$ GeV/$c^{2}$.%In the mass region $M_{ee}<0.14$ GeV/$c^{2}$, we report the centrality dependent $v_{2}$ shown in the bottom panel of Fig.3.

\begin{figure}[h]
\centering
%\begin{minipage}{12pc}
\includegraphics[width=18pc]{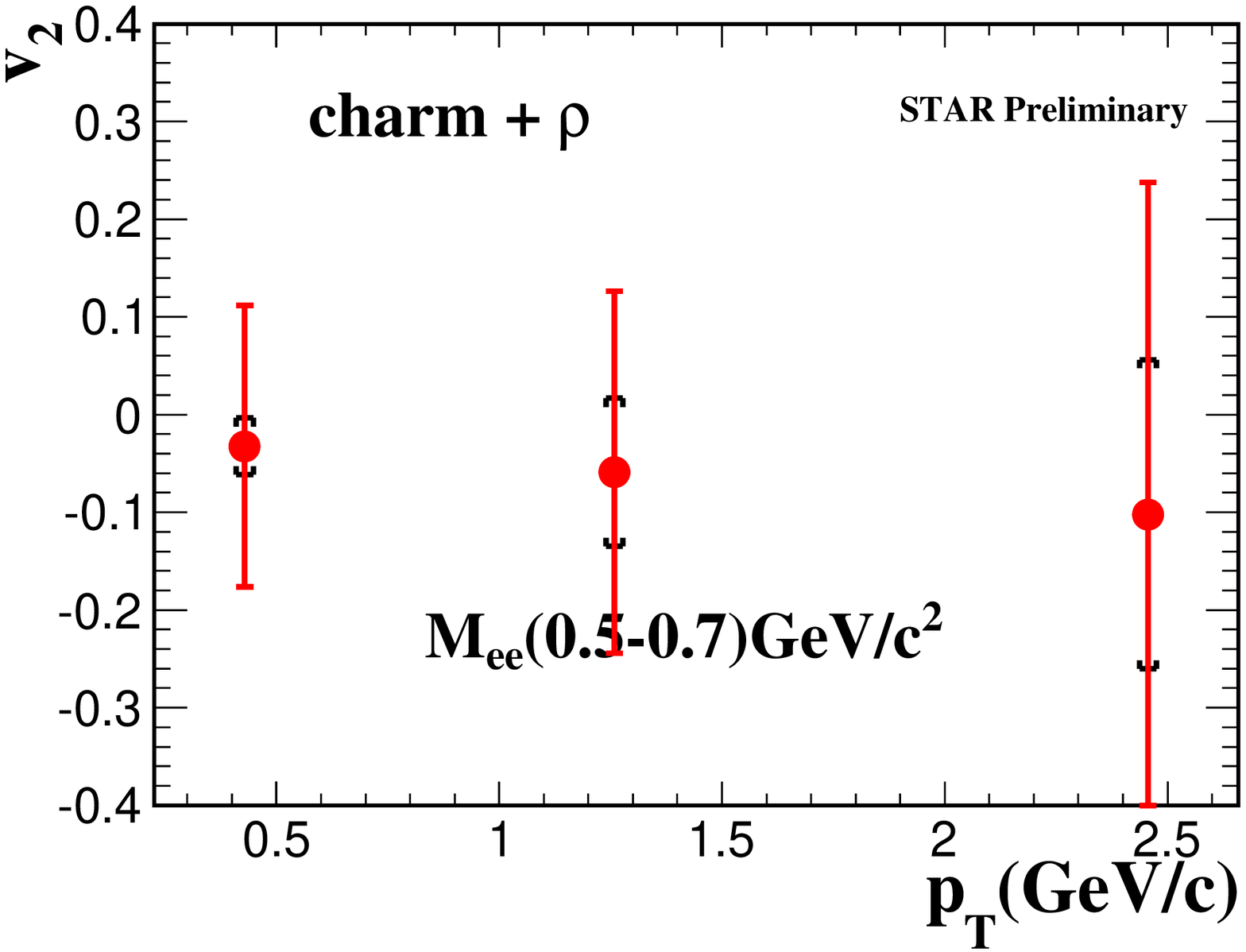}
%\end{minipage}\hspace{3pc}
%\begin{minipage}{12pc}
\includegraphics[width=18pc]{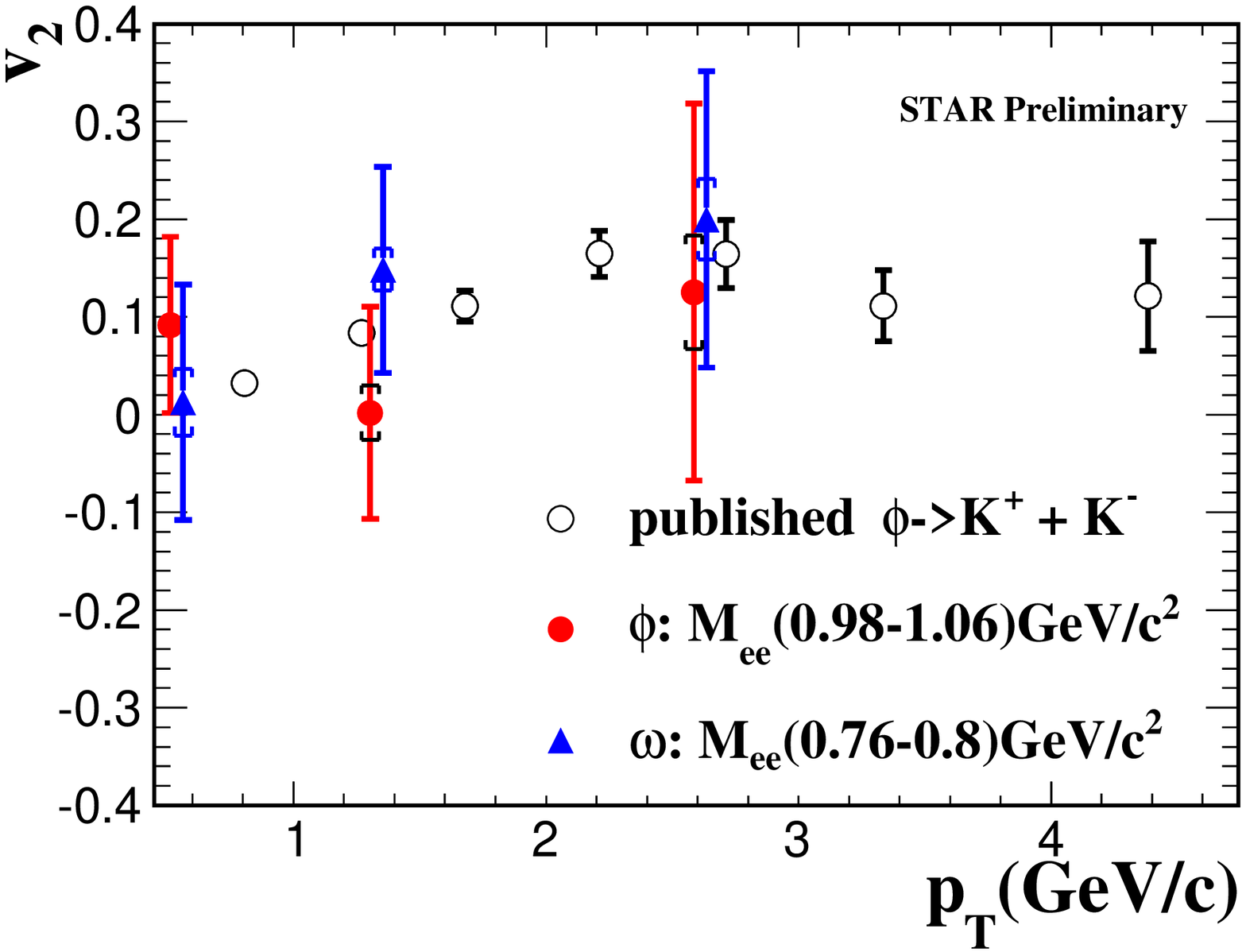}

\includegraphics[width=23pc]{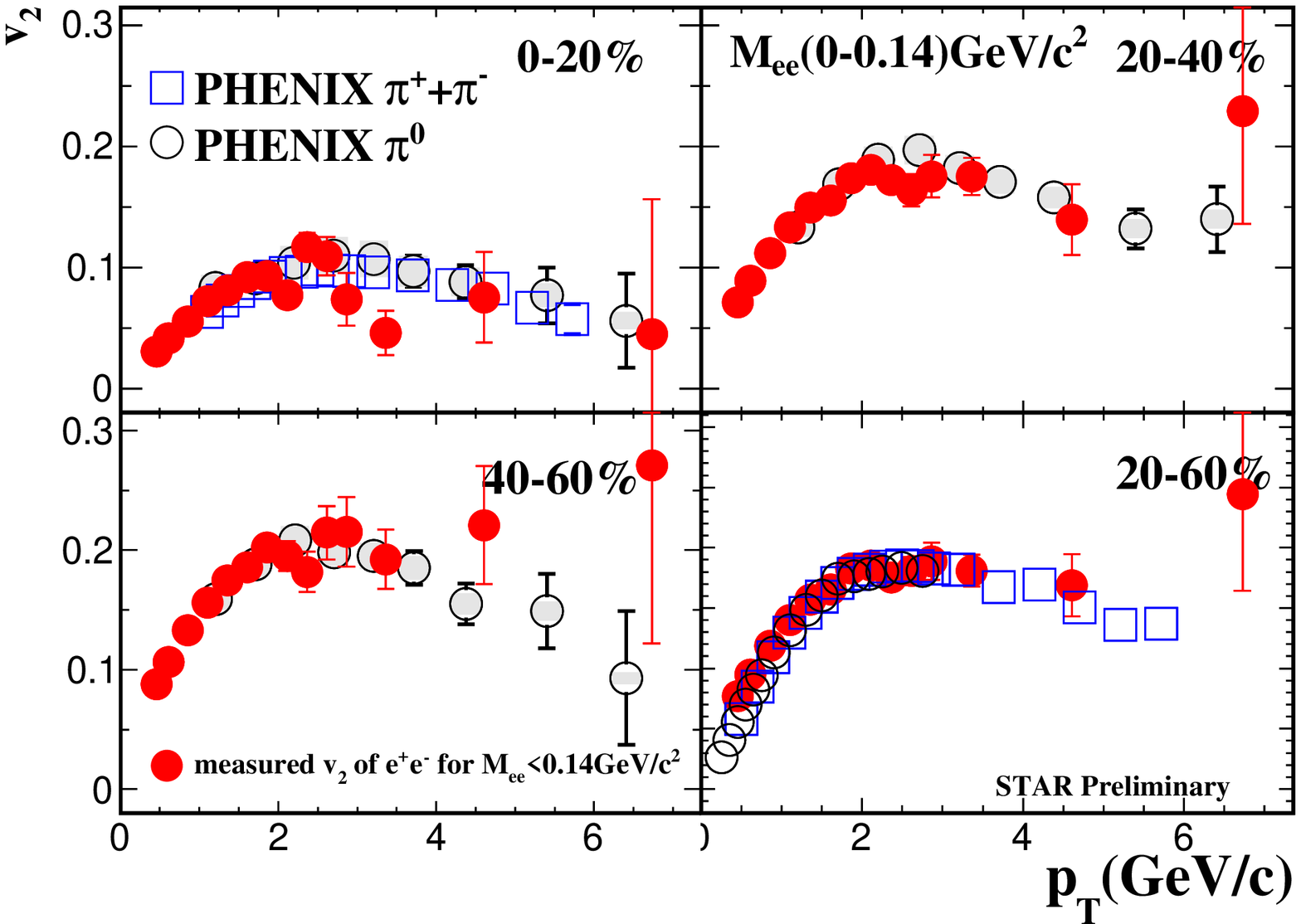}
\caption{\label{label}(upper-left panel) The di-electron $v_{2}$ as a function of $p_{T}$ in minimum bias Au+Au collisions at $\sqrt{s_{NN}}$=200 GeV for $0.5<M_{ee}<0.7$ GeV/$c^{2}$. (upper-right panel) The di-electron $v_{2}$ as a function of $p_{T}$ for $0.76<M_{ee}<0.8$ GeV/$c^{2}$  and for $0.96<M_{ee}<1.06$ GeV/$c^{2}$. The measured $v_{2}$ of $\phi$ meson through the $K^{+}K^{-}$ decay is also shown \cite{bibitem19}. The error bars and boxes represent statistical and systematic uncertainties, respectively. The systematic errors are estimated by changing track quality cuts. (bottom panel) The di-electron $v_{2}$ as a function of $p_{T}$ in different centralities for $M_{ee}<0.14$ GeV/$c^{2}$. The charged pion $v_{2}$ \cite{bibitem18} and neutral pion $v_{2}$ \cite{bibitem17} are also shown.}
\end{figure}
\section{Summary}

 Measurements of di-electron $v_{2}$ are presented as a function of $M_{ee}$ and $p_{T}$ from low to intermediate mass region in minimum-bias Au+Au $\sqrt{s_{NN}}$=200 GeV collisions. The elliptic flow of di-electron at $M_{ee}<0.14$ GeV/$c^{2}$ and $0.14<M_{ee}<0.3$ GeV/$c^{2}$ is in agreement with the expectations from previous measurements. For $M_{ee}<0.14$ GeV/$c^{2}$, the centrality dependent $v_{2}$ is also measured. In the future, more precise differential measurements will be obtained for di-electron $v_{2}$, as a result of the twice-larger dataset taken in the year 2011.

\end{document}